\documentclass{article}
\usepackage{amsmath, amssymb, amsfonts}
\title{On the source of the Kehagias - Sfetsos black hole}  
\author{Hristu Culetu, \\Ovidius University, Dept.of Physics, \\ Mamaia Avenue 124, 900527 Constanta, Romania, \\e-mail : hculetu@yahoo.com}

\begin{document}
\numberwithin{equation}{section}
\pagenumbering{arabic}
\maketitle
\newcommand{\fv}{\boldsymbol{f}}
\newcommand{\tv}{\boldsymbol{t}}
\newcommand{\gv}{\boldsymbol{g}}
\newcommand{\OV}{\boldsymbol{O}}
\newcommand{\wv}{\boldsymbol{w}}
\newcommand{\WV}{\boldsymbol{W}}
\newcommand{\NV}{\boldsymbol{N}}
\newcommand{\hv}{\boldsymbol{h}}
\newcommand{\yv}{\boldsymbol{y}}
\newcommand{\RE}{\textrm{Re}}
\newcommand{\IM}{\textrm{Im}}
\newcommand{\rot}{\textrm{rot}}
\newcommand{\dv}{\boldsymbol{d}}
\newcommand{\grad}{\textrm{grad}}
\newcommand{\Tr}{\textrm{Tr}}
\newcommand{\ua}{\uparrow}
\newcommand{\da}{\downarrow}
\newcommand{\ct}{\textrm{const}}
\newcommand{\xv}{\boldsymbol{x}}
\newcommand{\mv}{\boldsymbol{m}}
\newcommand{\rv}{\boldsymbol{r}}
\newcommand{\kv}{\boldsymbol{k}}
\newcommand{\VE}{\boldsymbol{V}}
\newcommand{\sv}{\boldsymbol{s}}
\newcommand{\RV}{\boldsymbol{R}}
\newcommand{\pv}{\boldsymbol{p}}
\newcommand{\PV}{\boldsymbol{P}}
\newcommand{\EV}{\boldsymbol{E}}
\newcommand{\DV}{\boldsymbol{D}}
\newcommand{\BV}{\boldsymbol{B}}
\newcommand{\HV}{\boldsymbol{H}}
\newcommand{\MV}{\boldsymbol{M}}
\newcommand{\be}{\begin{equation}}
\newcommand{\ee}{\end{equation}}
\newcommand{\ba}{\begin{eqnarray}}
\newcommand{\ea}{\end{eqnarray}}
\newcommand{\bq}{\begin{eqnarray*}}
\newcommand{\eq}{\end{eqnarray*}}
\newcommand{\pa}{\partial}
\newcommand{\f}{\frac}
\newcommand{\FV}{\boldsymbol{F}}
\newcommand{\ve}{\boldsymbol{v}}
\newcommand{\AV}{\boldsymbol{A}}
\newcommand{\jv}{\boldsymbol{j}}
\newcommand{\LV}{\boldsymbol{L}}
\newcommand{\SV}{\boldsymbol{S}}
\newcommand{\av}{\boldsymbol{a}}
\newcommand{\qv}{\boldsymbol{q}}
\newcommand{\QV}{\boldsymbol{Q}}
\newcommand{\ev}{\boldsymbol{e}}
\newcommand{\uv}{\boldsymbol{u}}
\newcommand{\KV}{\boldsymbol{K}}
\newcommand{\ro}{\boldsymbol{\rho}}
\newcommand{\si}{\boldsymbol{\sigma}}
\newcommand{\thv}{\boldsymbol{\theta}}
\newcommand{\bv}{\boldsymbol{b}}
\newcommand{\JV}{\boldsymbol{J}}
\newcommand{\nv}{\boldsymbol{n}}
\newcommand{\lv}{\boldsymbol{l}}
\newcommand{\om}{\boldsymbol{\omega}}
\newcommand{\Om}{\boldsymbol{\Omega}}
\newcommand{\Piv}{\boldsymbol{\Pi}}
\newcommand{\UV}{\boldsymbol{U}}
\newcommand{\iv}{\boldsymbol{i}}
\newcommand{\nuv}{\boldsymbol{\nu}}
\newcommand{\muv}{\boldsymbol{\mu}}
\newcommand{\lm}{\boldsymbol{\lambda}}
\newcommand{\Lm}{\boldsymbol{\Lambda}}
\newcommand{\opsi}{\overline{\psi}}
\renewcommand{\tan}{\textrm{tg}}
\renewcommand{\cot}{\textrm{ctg}}
\renewcommand{\sinh}{\textrm{sh}}
\renewcommand{\cosh}{\textrm{ch}}
\renewcommand{\tanh}{\textrm{th}}
\renewcommand{\coth}{\textrm{cth}}

\begin{abstract}
By assuming that the Kehagias-Sfetsos black hole is an exact solution of the standard Einstein equations, we investigate the properties of its source that generates the curvature. The anisotropic fluid has $p_{r} = - \rho$ as equation of state and fulfills the WEC and NEC. The gravitational field is repulsive inside the horizon and attractive outside, becoming of Schwarzschild type at large distances. The Misner-Sharp energy equals the black hole mass asymptotically.\\

 \end{abstract}
 
 \section{Introduction}
 An UV power counting renormalisable non-relativistic theory of gravity was proposed by Horava \cite{PH}. One may expect that its classical solutions are different from those of General Relativity (GR) at small scales, providing a mechanism to get rid of singularities \cite{CGS}. In addition, the theory does not exhibit the full 4-dimensional diffeomorphism invariance at large distances and deviations from GR could appear.
 
 In the framework of the Horava model, Kehagias and Sfetsos (KS) \cite{KS} found an asymptotically-flat spherically-symmetric spacetime which leads to the usual behavior of the Schwarzschild black hole (BH) at large distances from the source. However, their metric is singular at $r = 0$, even though the divergence is milder compared to Schwarzschild's. Setare and Momeni \cite{SM1, SM2} studied the geodesic stability and the spectrum of the entropy/area for the KS black hole in Horava-Lifshitz gravity via quasi-normal modes approach. 
 
 Recently Kim and Park \cite{KP} showed that there is an exact classical solution - a black wormhole - in the Horava quantum gravity model based on different scaling dimensions between space and time coordinates \cite{PH}. The solution depends on two quantities $M$ and $b$ where $M$ is the ADM mass and $b$ is a coupling constant of the theory (the positive IR parameter $b$ measures the strength of higher derivative corrections). Moreover, there is some repulsive interaction at short distances, the derivative of the metric coefficient diverges at $r = 0$ and we have a curvature singularity at the origin \cite{KP}. For some relation between the parameters $M$ and $b$ the BH becomes extremal and its horizon coincides with the wormhole throat. They so obtained a static black wormhole which is regular and interpolates between the BH state and the wormhole state through the coincidence state of an extremal BH \cite{KP}. 
 
 The purpose of this paper is to consider the KS metric (its extremal version) as a solution of the standard Einstein's equation and to find what stress tensor is needed on its r.h.s. for that geometry to be an exact solution. We reached the conclusion that the source represents an anisotropic fluid with $p_{r} = -\rho <0$, where $p_{r}$ is the radial pressure of the fluid and $\rho$ is its energy density. We also found that the gravitational field of the extremal BH is repulsive inside the horizon and the proper acceleration is equal to $1/\sqrt{3}m$ on the horizon, where $m$ is the BH mass. The quasilocal Misner-Sharp mass equals $m$ asymptotically. 
 
 Throughout the paper we use geometrical units $G = c = 1$.
 
 \section{Kehagias-Sfetsos spacetime} 
 KS studied the following asymptotically flat BH metric
   \begin{equation}
  ds^{2} = - f(r) dt^{2} + \frac{ dr^{2}}{f(r)} + r^{2} d \Omega^{2}, 
 \label{2.1}
 \end{equation}
in the framework of the Horava theory, where $m$ is for the time being an integration constant, $b$ is a positive constant related to the coupling constant of the theory, $d \Omega^{2}$ stands for the metric on the unit 2-sphere and
   \begin{equation}
	f(r) = 1+ br^{2}- \sqrt{r(b^{2}r^{3} + 4bm)}.
 \label{2.2}
 \end{equation}
The geometry (2.1) acquires the usual Schwarzschild behavior
   \begin{equation}
	f(r) \approx 1 - \frac{2m}{r} + O(1/r^{4})
 \label{2.3}
 \end{equation}
for $r >> (\frac{r}{b})^{1/3}$. The KS line element is singular at $r = 0$ and has two horizon at \cite{KS}
   \begin{equation}
	r_{\pm} = m\left(1 \pm \sqrt{1 - \frac{1}{2bm^{2}}}\right),
 \label{2.4}
 \end{equation}
with $2bm^{2} \geq 1$.

To simplify the calculations, we are interested in the extremal case $2bm^{2} = 1$ when the two event horizon coincide at $r = m$ and the model depends on one parameter only. We see from (2.1) and (2.2) that the derivative
   \begin{equation}
	f'(r) = 2b \left(r - \frac{br^{3} + m}{\sqrt{b^{2}r^{4} + 4bmr}}\right)
 \label{2.5}
 \end{equation}
vanishes at $r_{0} = (\frac{m}{2b})^{1/3}$ which is a stationary point of the function $f(r)$. In the extremal case $r_{0} = m$ and the plot of $f(r)$ versus $r$ touches the abscissa at the extremal point \cite{KP}. 

\section{Kinematic considerations}
Let us now take a static observer in the extremal KS spacetime, characterized by the velocity vector field
   \begin{equation}
	u^{b} = \left(\frac{x \sqrt{2}}{\sqrt{1 + 2x^{2} - \sqrt{1 + 8x^{3}}}}, 0, 0, 0 \right),
 \label{3.1}
 \end{equation}
where $x = m/r$ and $u^{b}u_{b} = -1$ ($b$ labels here the $t, r, \theta, \phi$). From (3.1) we obtain the acceleration 4-vector of the static observer
   \begin{equation}
	a^{b} = \left(0, \frac{\sqrt{1 + 8x^{3}} - 2x^{3} - 1}{2mx \sqrt{1 + 8x^{3}}}, 0, 0 \right).
 \label{3.2}
 \end{equation}
From (3.2) we observe that $a^{r}$ vanishes at the horizon $r = m ~(x = 1)$ and at $x = 0$ (we have actually to take the limit $x \rightarrow 0$ because the function $a^{r}(x)$ is not defined at $x = 0$). In addition, the gravitational field is repulsive for $x > 1 ~(r < m)$ and attractive for $x < 1 ~(r > m)$. That reminds us of a similar property encountered previously (see \cite{HC1} for the study of the radial acceleration of a static observer for a regular modified version of the Schwarzschild geometry). Another similarity with \cite{HC1} comes from the fact that $f(r)$ has no a signature flip when the horizon $r = m$ is crossed ($f(r)$ from (2.2) is positive everywhere, as in \cite{HC1}). Moreover, even the plots of $f(r)$ from (2.2) (see \cite{KP}) and of $f(r) = 1 - (2m/r) exp(-2m/er)$ from \cite{HC1} are very similar, with the same boundary values. Again from (3.2) we obtain for $x << 1$ (or $r >> m)$ that
 \begin{equation}
	a^{r} \approx \frac{x^{2}}{m} (1 + 4x^{3}) = \frac{m}{r^{2}} (1 + \frac{4m^{3}}{r^{3}}),
 \label{3.3}
 \end{equation}
where we recognize the 1st term as the Newtonian term.

The proper acceleration is given by
  \begin{equation}
 a \equiv  \sqrt{ a^{b}a_{b}} = \frac{\sqrt{2}~ x^{2} (1 + x + x^{2})\sqrt{1 + 2x^{2} + \sqrt{1 + 8x^{3}}}}{m (1 + 2x^{3} + \sqrt{1 + 8x^{3}}) \sqrt{1 + 8x^{3}}},       
 \label{3.4}
 \end{equation} 
with $a = 1/\sqrt{3}m$ on the horizon $x = 1$, $a = 0$ asymptotically ($x \rightarrow 0$) and $a \rightarrow \infty$ at the origin $r = 0 ~(x \rightarrow \infty)$. However, the surface gravity 
 \begin{equation}
\kappa =  \sqrt{ a^{b}a_{b}}~ \sqrt{-g_{tt}}|_{r = m} = 0,
 \label{3.5}
 \end{equation}
since the BH is extremal. In other words, the Hawking temperature is vanishing (no Hawking radiation) and we are left with a ''frozen'' horizon \cite{BR}.

In spite of the fact that the curvature invariants $R^{a}_{~a}$ and the Kretschmann scalar $K$ are divergent at the origin, they are finite on the horizon, with $R^{a}_{~a} = 4/3m^{2}$ and $K = 880/243m^{4}$ at $r = m$.

\section{Anisotropic stress tensor}
For the metric (2.1) with $f(r)$ from (2.2) to be a solution of Einstein's equation $G_{ab} = 8\pi T_{ab}$, we must have on its r.h.s. the following expressions for the nonzero components of the energy-momentum tensor of the source
 \begin{equation}
   \begin{split}
  -8\pi T^{t}_{~t} = 8\pi \rho = \frac{3}{2m^{2}} \left(\frac{1 + 4x^{3}}{\sqrt{1 + 8x^{3}}} - 1\right), ~~~ 8\pi T^{\theta}_{~\theta} = 8\pi T^{\phi}_{~\phi} = \\ 8\pi p_{\theta} = 8\pi p_{\phi} = \frac{3}{2m^{2}} \left(1 - \frac{8x^{6} + 12x^{3} + 1}{(1 + 8x^{3}) \sqrt{1 + 8x^{3}}}\right) .
  \end{split}
\label{4.1}
\end{equation}
A simpler way to compute the radial component $G^{r}_{~r}$ of the Einstein tensor is to get it from the scalar curvature
 \begin{equation}
R^{a}_{~a} = \frac{6}{m^{2}} \left[\frac{(1 + 2x^{3})(1 + 10x^{3})}{(1 + 8x^{3}) \sqrt{1 + 8x^{3}}  } -1\right],
 \label{4.2}
 \end{equation}
whence
 \begin{equation}
  G^{r}_{~r} = - G^{t}_{~t} - 2 G^{\theta}_{~\theta} - R^{a}_{~a} = G^{t}_{~t},
 \label{4.3}
 \end{equation}
so that $p_{r} = -\rho$, as for the dark energy. 

All the components of the stress tensor are vanishing asymptotically and are divergent at $r = 0 ~(x \rightarrow \infty)$. Nevertheless, they are finite at the horizon where we have
 \begin{equation}
\rho = - p_{r} = 3 p_{\theta} = 3 p_{\phi} = \frac{1}{8\pi m^{2}}
 \label{4.4}
 \end{equation}
where $p_{\theta}$ and $p_{\phi}$ are the transversal pressures. As a function of $x$, the energy density is a monotonic increasingly function as may be seen from
 \begin{equation}
\rho'(x) = \frac{72x^{5}}{m^{2}(1 + 8x^{3}) \sqrt{1 + 8x^{3}}} > 0,
 \label{4.5}
 \end{equation}
which diverges asymptotically ($x \rightarrow \infty$).

As far as the energy conditions are concerned, we see from (4.1) and (4.3) that the WEC and NEC are always satisfied but the SEC is not obeyed when $\rho + p_{r} + 2p_{\theta} < 0$, namely for $r < (3\sqrt{3} - 5)^{1/3} m \approx 0.575 m$. The DEC is also not observed when $\rho < |p_{\theta}|$, i.e. for $r > (3\sqrt{6} - 2)^{1/3} m \approx 1.74 m$. Hence, all the energy conditions are satisfied only in the narrow range $0.575 m \leq r \leq 1.74 m$, where the horizon $r = m$ is located.

\section{Misner - Sharp energy}
Our next task is to evaluate the quasilocal Misner - Sharp  mass $E(r)$ \cite{CG, FMN, NY, HC2} from
 \begin{equation}
1 - \frac{2E(r)}{R} = g^{ab}\partial_{a}R \partial_{b}R,
 \label{5.1}
 \end{equation}
where $R$ is the areal radius (here we have $R = r$). One obtains
 \begin{equation}
E(x) = \frac{m}{4x^{3}} (\sqrt{1 + 8x^{3}} - 1),
 \label{5.2}
 \end{equation}
in terms of $x = m/r$. We have $E = 0$ at $r = 0$ and $E = m$ (the ADM mass) at infinity, after the limiting process. This last property justifies the choice of $m$ as the BH mass, including the energy of the anisotropic fluid. It is worth noting that because of the lack of a signature flip when the horizon is crossed, the spacetime (2.1)-(2.2) is valid both inside and outside the BH. In addition, $dE(r)/dr > 0$ so that $E(r)$ is a monotonically increasing function of $r$. We have also $E(1) = m/2$. That means half of the BH mass is located outside the horizon.

\section{Conclusions}  
The extremal Kehagias-Sfetsos black hole has been investigated in this paper. Even though their metric was obtained in the framework of the Horava theory, we took it as an exact solution of the conventional Einstein equations and looked for the stress tensor generating it. We found it corresponds to an anisotropic fluid with $p_{r} = - \rho$ as the equation of state, as for the dark energy. Although the mathematical expressions of the physical quantities implied are complicate, they exhibit simple and useful properties, being finite on the BH horizon $r = m$. Because of the negative radial pressure the gravitational field is repulsive inside the horizon. We also found that the quasilocal Misner-Sharp energy equals the BH mass asymptotically.

\end{document}